\documentclass[a4paper,10pt]{article}
\usepackage[T1]{fontenc}
\usepackage[utf8]{inputenc}
\usepackage[italian,english]{babel}
\usepackage[autostyle,italian=guillemets]{csquotes}
\usepackage{multicol}
\usepackage{microtype}
\usepackage{guit}
\usepackage{booktabs}
\usepackage{graphicx}
\usepackage{mathptmx}
\usepackage{amsfonts}
\usepackage{amssymb}
\usepackage{amsmath}
\usepackage{ifpdf}
\usepackage{dcolumn}
\usepackage{textcomp}
\usepackage{tabularx,array}
\usepackage{ulem}
\usepackage{placeins}
\usepackage{epstopdf}
\usepackage{etex}
\usepackage{rotating}
\date{}

\title{PAN.R.C.- A New Level 3 Biosafety and Astrobiology Laboratory in Pieve a Nievole (PT)}
\author{Tasselli, D. \\ TS Corporation Srl - Astronomy and Astrophysics Department\\ Via Rugantino, 71, 00169 Roma RM - Italy \\ E-mail:diego.tasselli@tscorporation.org \\ \\ Ferrara, G. \\ TS Corporation Srl - Biology Department\\ Via Cavalcanti, 10, 51010 Massa e Cozzile PT - Italy \\ E-mail:giuseppina.ferrara@tscorporation.org \\ \\ Ricci, S.\\ TS Corporation Srl - Meteorogical and Climatic Change Department\\
Via Rugantino, 71, 00169 Roma RM - Italy \\ E-mail:stefano.ricci@tscorporation.org \\ \\ Bianchi, P.\\ TS Corporation Srl - Geology and Geophysics Department \\ Via Rugantino, 71, 00169 Roma RM - Italy\\ E-mail:pamela.bianchi@tscorporation.org}
\begin{document}
	
\maketitle
\begin{abstract}
\normalsize We report our proposal for the establishment of a biocontainment and astrobiology laboratory in a strategic area of Pieve a Nievole (PT) at 28 mt above sea level - to face the lack of biological and astrobiological research centers and all the social, economic and cultural consequences that this project implicate. The structure will be built under the Horizon 2020 work program 2018-2020 - European Research Infrastructures (including e-Infrastructures), and will enable the development of major research project.
\end{abstract}

{\footnotesize This paper was prepared with the \LaTeX} \\
\begin{multicols}%
{2}

\section{\normalsize Introduction}
The environment that surrounds us is populated by biological agents such as bacteria, viruses or fungi. Some of these inhabit the surface of our skin or even form part of our intestinal flora. \\
Not all of them are harmful, in fact, many of these microorganisms allow us a better life. However, a small part of these organisms can get into our bodies and cause problems.\\
By the term “biological risk” we mean every kind of risk deriving from the direct or indirect contact with these agents that can cause disease.\\ 
To date, biological risk is managed though the classification of biological agents in four groups, based on the effects they cause on healthy people, as demonstrated by {\itshape Figure 1}.\\ \\
\includegraphics[width=0.99\linewidth]{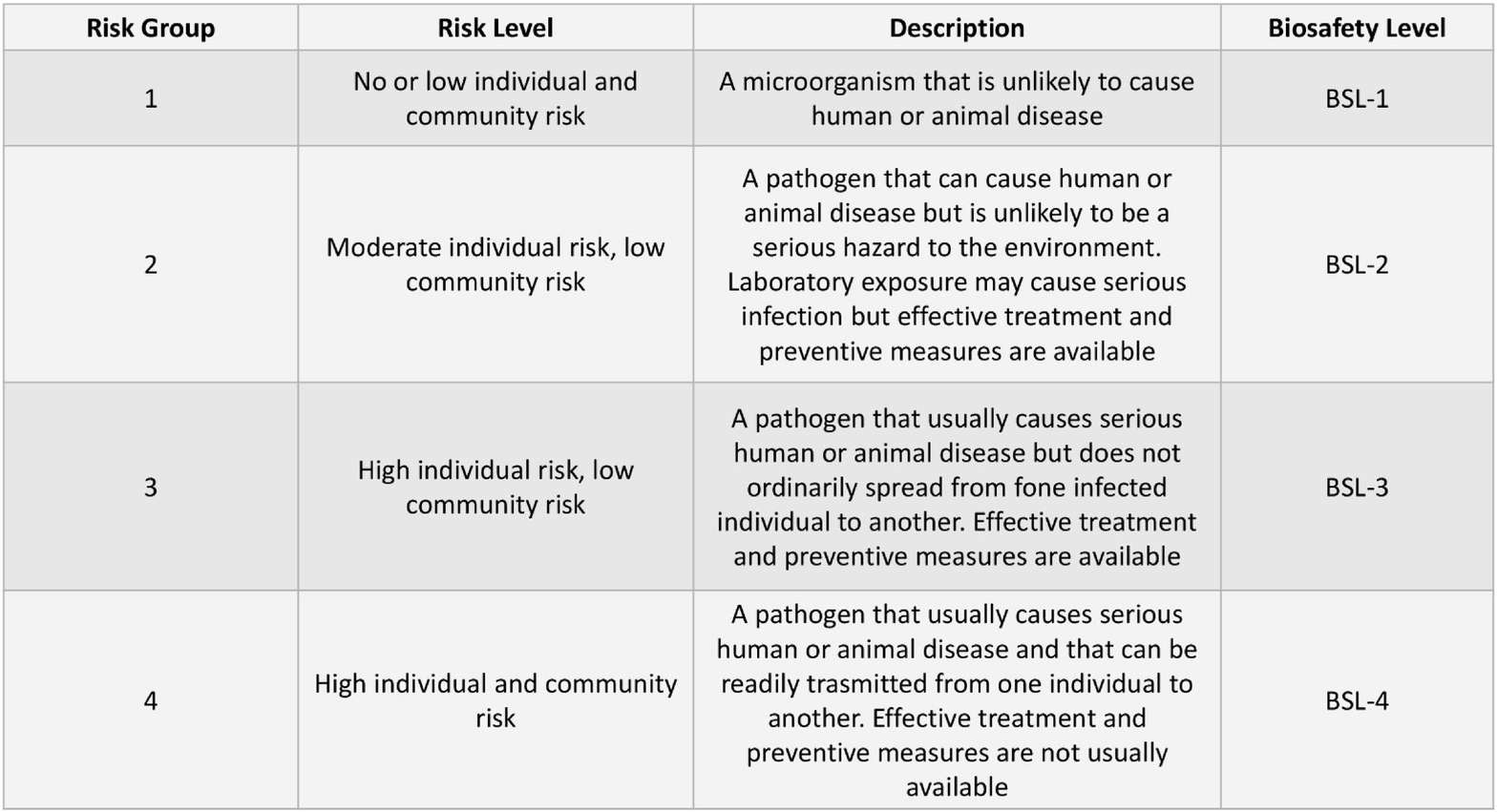}\\
\textbf{{\scriptsize Fig.1 - Classification of microorganisms by risk groups.}} \\ \\
According to this classification, four specific containment criteria there have been identified, to which the diagnostic or research laboratories must refer.
 \section{\normalsize State of art in Italian research centers}
To date, in Italy there are only four research centers with a level 3 biosafety laboratory. The lack of facilities on the Italian territory represents a serious problem, because of the return of previously eradicated endemic disease and the onset of new viral and bacterial illness due to globalization, overcrowding and immigration.\\
Until now, any company, university or researcher, who wants to have part in a project that involves the use of biological agents from second or third group, must ask permission and wait for the technical time to use the currently available facilities. \\ Moreover, the participation of students from other countries in both the Euro and non-Euro zone is reduced at the moment, with a consequent reduced return in productivity, efficiency, economy and internationalization.

\section{\normalsize Why this place}
Pieve a Nievole (PT) is deemed to be the best place in the district of Montecatini Terme (PT) and in the Valdinievole area. \\ \\
% Table generated by Excel2LaTeX from sheet 'Dati Geografici'
\begin{tabular}{|ll|}
	\hline
	\scriptsize Latitude & \scriptsize 43°52'21"72 N \\
	\hline
	\scriptsize Longitude & \scriptsize 10°48'9"00 E \\
	\hline
	\scriptsize Share & \scriptsize 28 mt. s.l.m. \\
	\hline
\end{tabular}  
\\ \\
\textbf{{\scriptsize Table 1. Pieve a Nievole (PT) coordinates.}} \\ \\
Thanks to its meteorological and geological conditions, this area is perfect for building and insediation a research laboratory, thank to the average weather conditions in certain times of the year (Table 2).\\
The main advantages of Pieve a Nievole (PT) are summarized as follows:
\begin{itemize}
\item exceptional air stability; 
\item continuous research activity;
\item proximity to two airport: Florence to 60 Km and Pisa to 40 km;
\item adjacency to the florence-sea highway
\end{itemize}
\subsection{\normalsize Meteo-Clima Data}
Preliminary tests on the site were performed with good results. \\
Data on cloud cover to Pieve a Nievole (PT) has acquired automatically, showed clear skies during most of the winter season, while the average value of maximum wind speed is watched in Table 2.
% Table generated by Excel2LaTeX from sheet 'Medie Climatica'
\\ \\
\begin{tabular}{|cccccc|}
	\hline
	{\bf \scriptsize Mouth} & {\bf \scriptsize T min} & {\bf \scriptsize T max} & {\bf \scriptsize Prec.} & {\bf \scriptsize Umidity} & {\bf \scriptsize Wind} \\ &{\scriptsize °C} &{\scriptsize °C}&{\scriptsize mm}&{\scriptsize \%}&{\scriptsize Km/h}
	\\
	\hline
	\multicolumn{ 6}{|c|}{} \\
	\hline
	\scriptsize Gen &   \scriptsize 2  & \scriptsize 11  &  \scriptsize  74  &  \scriptsize  75 & \scriptsize   E 9 \\
	\hline
	\scriptsize Feb &   \scriptsize  3  &  \scriptsize 12  &  \scriptsize  70  & \scriptsize  71 & \scriptsize  E 9 \\
	\hline
	\scriptsize Mar & \scriptsize   5  & \scriptsize 15  & \scriptsize  77  & \scriptsize  70 & \scriptsize  W 16 \\
	\hline
	\scriptsize Apr & \scriptsize   7  & \scriptsize  18  & \scriptsize  80  & \scriptsize  72 & \scriptsize  W 16 \\
	\hline
	\scriptsize May & \scriptsize  11  & \scriptsize 22  &  \scriptsize  61  & \scriptsize  72 & \scriptsize  W 16 \\
	\hline
	\scriptsize Jun & \scriptsize 14  & \scriptsize 26  & \scriptsize  43  & \scriptsize 70 & \scriptsize W 16 \\
	\hline
	\scriptsize Jul &\scriptsize  17  & \scriptsize 29  & \scriptsize 24  & \scriptsize 67 & \scriptsize  W 16 \\
	\hline
	\scriptsize Aug & \scriptsize  17  & \scriptsize  29  & \scriptsize  57  & \scriptsize  68 & \scriptsize  W 16 \\
	\hline
	\scriptsize Sep &\scriptsize 14  & \scriptsize 26  & \scriptsize 88  & \scriptsize 71 & \scriptsize  W 16 \\
	\hline
	\scriptsize Oct &\scriptsize  11  & \scriptsize 21  & \scriptsize 120  &\scriptsize 72 & \scriptsize  W 9 \\
	\hline
	\scriptsize Nov &\scriptsize  6  & \scriptsize 16  & \scriptsize 122  & \scriptsize 74 & \scriptsize E 9  \\
	\hline
	\scriptsize Dec &\scriptsize  3  & \scriptsize 12  &\scriptsize 85  &\scriptsize 76 & \scriptsize E 9  \\
	\hline
\end{tabular}  
\\ \\
\textbf{{\scriptsize Table 2: Average weather conditions over the past 30 years.}}
\subsection{\normalsize Geological data}
From a geological point of view, you can not look at the area of the Pieve a Nievole town, without inserting it into a wider contest that includes the whole surrounding flat area.\\
The area affected by the project is a flat area characterized by the presence of the large {\itshape "swamp of Fucecchio"} whose drainage occurred in various historical epoche.\\
This type of area is frequent in Tuscany, and is a phenomenon resulting from the genetic phenomenon due to the reduction of the comprehensive efforts that have originated in the Apennine chain. \\
The beginning of the evolution of the area can be summarized as follows:
\begin{itemize}
	\item \textbf{Middle-upper Pliocene (from 4 to 2 Million Years ago)}\\
	If Italy is on a larger scale conceived as already defined in the main orographic characteristics, the situation seen in the area covered by the present study is different, as this area consists of a slope with a slight slope crossed by rivers that descend from the north.\\
\end{itemize}
\begin{itemize}
\item \textbf{Lower Pleistocene (Villafranchiano Superiore from 2 to 0.7 million years ago)}\\
This period is characterized by the progressive lifting of the area due to a progressive subsidence, linked to the phenomena of migration to the east of the comprehensive efforts responsible for the birth of the Apennines. And the consequence of these events was the birth of a lake in which the paleo rivers brought sediments, while the ground due to the effect of subsidenze continued to lower until it touches 400 meters below sea level. At the end of this period, probably due to a glacial event, the sea level was lowered, until it came to an offset of the coast, not unlike the current configuration.\\
In this period also the lakes present disappeared, and specifically in the area affected by the project, this phenomenon due to the lower subsidence compared to the sedimentary contribution, determined a reduction rather than a disappearance of the lake basin of Pescia-Empoli In this period, the residual \textit{"Fucecchio lake area"} turned into a swamp.
\end{itemize}	
\begin{itemize}
	\item \textbf{ Middle-upper Pleistocene (from 700,000 to 8,800 years ago)}\\
	In this period the area still affected by elevation, assumed the current configuration. There were no substantial orographic changes except for a drift of the swamp that occupied the eastern side.	
\end{itemize}
\begin{itemize}
	\item \textbf{ Holocene (from 8,800 years to current date)}\\
	In this perimeter the variations are almost all of an anthropic nature. In fact, during the \textit{"Leopoldino period"} we can see the reclamation of the Fucecchio marsh which has reduced its extinction and regulated the waters.
\end{itemize}
\section{\normalsize PAN.R.C. in summer}
The research facility will consist of two area on either side of the structure with a central area of calculation and handling of electronic instrumentation. \\ \\
\includegraphics[width=0.99\linewidth]{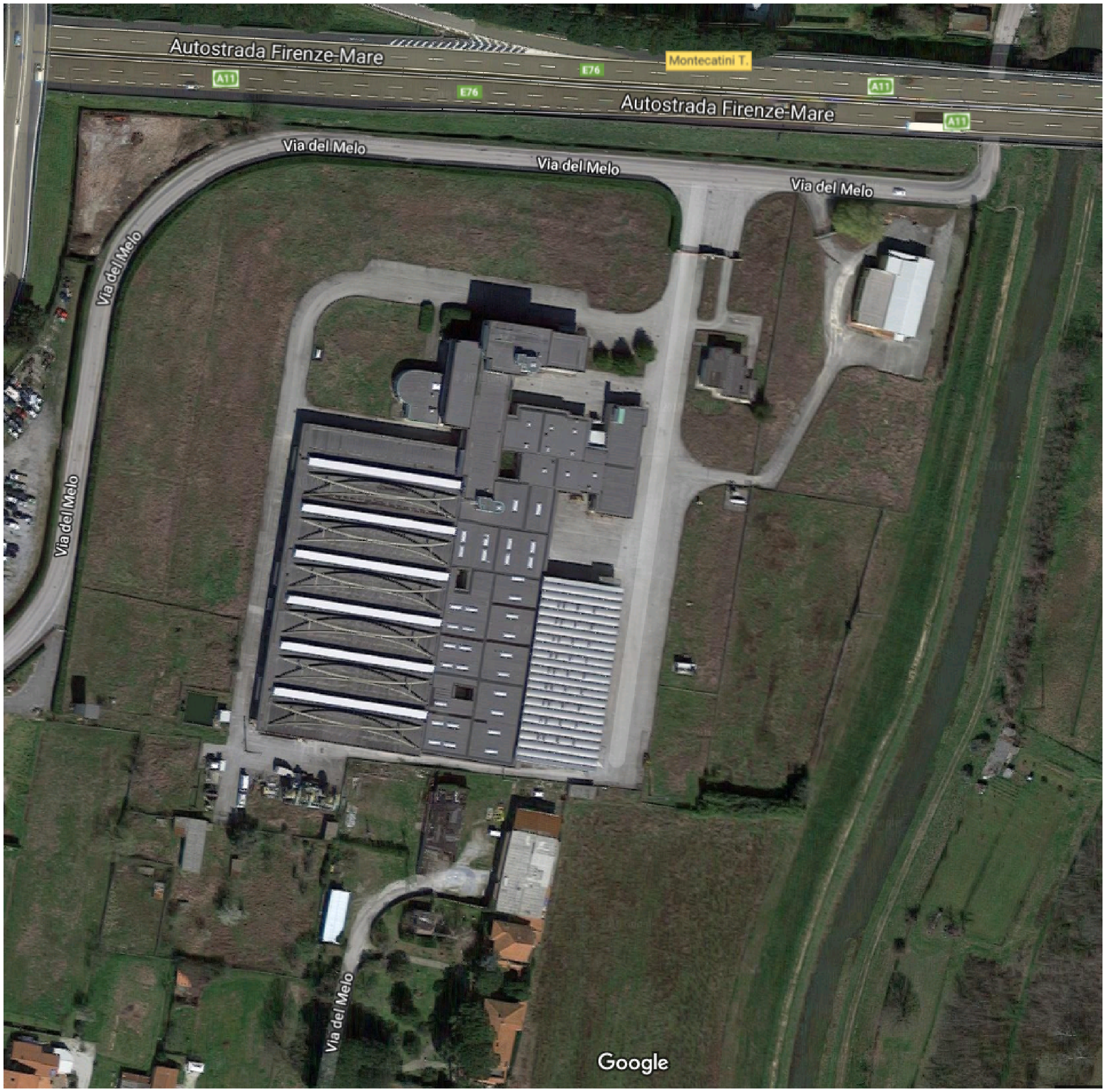}\\
\textbf{{\scriptsize Fig.3 - PAN.R.C. with Google Eart}} \cite{Google:ref1} \\ \\
This infrastructure must be correlated with a series of equipment that can guarantee not only the effectiveness of research, but also the maximum containment of the agents treated. 
\subsection{\normalsize BSL 3 }
A level 3 laboratory, specifically, is suitable for study, analysis and research of biological agents, such as bacteria, viruses and fungi  belonging to the second and third groups. Also continuous research activity in a list of uman patology. \\ For this reason, level 3 biosafety laboratories must be built in a strategic area with law-defined criteria.
\subsection{\normalsize Future perspective}
The addition of a new research facility in this area, together with those already present, would help to increase employment, expand the research area and the physical space reserved for it.\\ 
Cultural exchange with other countries would be favored part of the facility could be destined to host currently unavailable avant-garde instruments, which can also be used by other Italian research centers. \\
The presence of this facility will contribute to the growth of the scientific community of biologists and biotechnologists, as well as it will encourage the development of local companies.\\
A laboratory construction and maintenance, in fact, cannot be possible without the support of artisan and professional industries, from the hydraulic to the electrical, from the design and installation of ventilation systems to their management, from construction of automated doors for the access the laboratory to the maintenance of filters in air systems ({\itshape Figure~2}).\\
\includegraphics[width=1\linewidth]{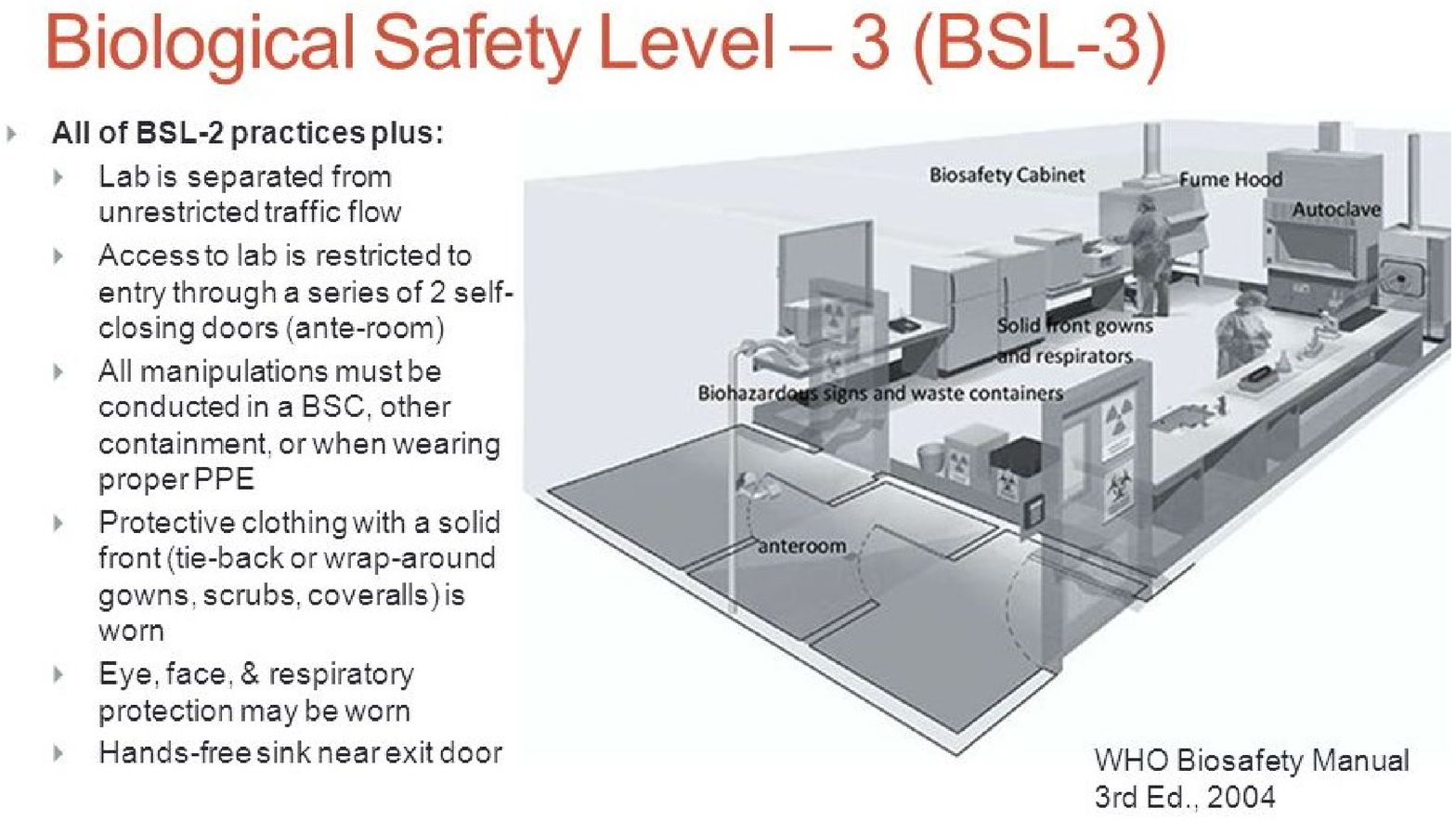}\\
\textbf{{\scriptsize Fig. 2 - A typical Level 3 Biosafety Laboratory}}\\ \\
An essential characteristics of research centers is the ability to create a network between members of scientific community, with the aim of expand it more and more. Sharing a research project and its results would set a multidisciplinary work and support its distribution and dissemination in several scientific fields beyond biology.\\
The availability of additional resources in these fields will encourage interaction between sciences and the development of significant projects, initiatives and discoveries.\\
The laboratory will collaborate with international universities and will certainly became an incentive for students who wants to approach the research or scientific analysis.

\section{\normalsize Conclusions}
In conclusion, the establishment of a new level 3 biosafety and astrobiology laboratory in Pieve a Nievole (PT) would be an added value to the Italian research, economy and society and would extend the international network interactions among scientists of European Countries. \\ The structure with its declared capacity will also be available for current and upcoming projects in biological, and expectations can not be anything but very good.

\section{\normalsize Acknowledgments}
We would like to thank Silvia Marcoccia for supporting with suggestions and presence during data analysis. \\The constructive comments are highly appreciated.
\end{multicols}


\clearpage
\begin{thebibliography}{99}
	\bibitem{Tasselli1:ref1}
	\textbf{Tasselli, D}, Investimento presso il Comune di Pieve a Nievole (PT), 2018
	\bibitem{Google:ref1}
	\textbf{Google Earth}, map of research center in Pieve a Nievole (PT), 2018
	\bibitem{Boccaletti:ref1}
	\textbf{Boccaletti, M. \& calamita, F. et all}, "Evoluzione dell'Appennino Tosco-Umbro-Marchigiano durante il Neogene", Giornale di Geologia, Sez.3, vol.48/1-2, 1986
	\bibitem{Capecchi:ref1} 
	\textbf{Capecchi, F. \& Pranzini, G.}, "Studi geologici e idrogeologici della pianura di Pistoia", Boll.Soc.Geol. It. 94, 1975
	\bibitem{Destefani:ref1}
	\textbf{De Stefani C.}, "I dintorni di Monsummano e Montecatini in Val di Nievole", Boll. Soc. Geol. It. 42-52, 1887
	\bibitem{Frazzuoli:ref1}
	\textbf{Frazzuoli, M. \& Maestrelli-Manetti O. }, "I nuclei Mesozoici di Monsummano, Montecatini Terme e Marliana (provincia di Pistoia)", 1973
	\bibitem{Ghelardoni:ref1}
	\textbf{Ghelardoni, R.}, "Spostamento dallo spartiacque dell'Appennino Settentrionale in corrispondenza di catture idrografiche", Atti Soc. Sc. Nat., Mem. Ser. A, 65, 1958
	\bibitem{Arrighi:ref1}
	\textbf{Arrighi, A. \& Bertogna, A. \& Naef, S.}, "Montalbano: geologia, flora, fauna, storia e arte", Cnsorzio Interprovinciale per il Montalbano, 1993
	\bibitem{Horizon2020:ref1} 
	\textbf {Horizon 2020 - Work Programme 2018-2020}, European research infrastructures (including e-Infrastructures), 2016.
	\bibitem{Devendra:ref1} 
	\textbf{Devendra T. Mourya et all.},"Establishment of Biosafety Level-3 (BSL~3) laboratory: Important criteria to consider while designing, constructing, commissioning \& operating the facility in Indian setting", 2014, Indian J Med Res.
		\bibitem{Callaway:ref1} 
	\textbf{Callaway E.},"Biosafety concerns for labs in the developing world", 2012, Nature
	\bibitem{Airespa:ref1}
	\textbf{Manuale di Biosicurezza nei laboratori},  Ed. Airespa, 2005
	\bibitem{ISO:ref1}
	\textbf{UNI EN ISO 15189:2007 – Medical Laboratories. Requirements for quality and competence – International Organization for Standardization}
	\bibitem{Adel:ref1}
	\textbf{Adel N.,Zaky},"Biosafety and biosecurity measures: management of biosafety level 3 facilities", 2010, International Journal of Antimicrobial Agents
	\bibitem{WHO:ref1}
	\textbf{Laboratory biosafety manual}, 2004, World Health Organization, Geneva
	\bibitem{EUDirective:ref1}
	\textbf{EU Directive 2000/54 – Protection of workers from risks related to exposure to biological agents at work}, 2000
	\bibitem{EBSA:ref1}
	\textbf{EBSA - Revision of the EU biological agents directive 2000/54/EC}, 2000
	\bibitem{DM593:ref1}
	\textbf{Art. 7 del D.M. n. 593, 8 agosto 2000}
	\bibitem{DM593:ref2}
	\textbf{Art. 12 del D.M. n. 593, 8 agosto 2000}
	\bibitem{DM593:ref3}
	\textbf{Art. 13 del D.M. n. 593, 8 agosto 2000}
	\bibitem{CC:ref1}
	\textbf{Libro I°Titolo secondo, del Codice Civile}
	\bibitem{Regcom:ref1}
	\textbf{Regolamento Comunale di Pieve a Nievole – PT}
\end{thebibliography}
\end{document}